\documentclass[11pt,draftcls,onecolumn]{IEEEtran}
\usepackage{graphicx,amssymb,amsmath,color}
\usepackage{subfigure}
\usepackage{url}
\usepackage[square, comma, sort&compress, numbers]{natbib}

\usepackage{multirow} 
\usepackage{geometry}
\title{A Survey on High-Speed Railway Communications: A Radio Resource Management Perspective}
\author{Shengfeng Xu, Gang Zhu, Bo Ai, Zhangdui Zhong\\State Key Laboratory of Rail Traffic Control and
Safety\\Beijing Jiaotong University, Beijing, P. R. China
\\Email:$\{\text{xsf1988, gzhu, boai, zhdzhong}\}$@bjtu.edu.cn}

\begin{document}

\maketitle
\begin{abstract}
High-speed railway (HSR) communications will become a key feature supported by intelligent transportation communication systems. The increasing demand for HSR communications leads to significant attention on the study of radio resource management (RRM), which enables efficient resource utilization and improved system performance. RRM design is a challenging problem due to heterogenous quality of service (QoS) requirements and dynamic characteristics of HSR wireless communications. The objective of this paper is to provide an overview on the key issues that arise in the RRM design for HSR wireless communications. A detailed description of HSR communication systems is first presented, followed by an introduction on HSR channel models and characteristics, which are vital to the cross-layer RRM design. Then we provide a literature survey on state-of-the-art RRM schemes for HSR wireless communications, with an in-depth discussion on various RRM aspects including admission control, mobility management, power control and resource allocation. Finally, this paper outlines the current challenges and open issues in the area of RRM design for HSR wireless communications.
\end{abstract}


\section{Introduction}
For the last two decades, intelligent transportation systems (ITS) have emerged as an efficient way of improving the performance of transportation systems.
As an essential element of ITS, high-speed railway (HSR) has been developed rapidly as a fast, convenient and green public transportation system and would become the future trend of railway transportation worldwide \cite{Ning-2011}.
For instance, a high speed rail plan has been outlined in America and the length of HSR lines in China will reach 18,000 km by 2020 \cite{Wang-2012}.
With the continuous construction of HSR in recent years, the issue of train operation safety has attracted more and more attention.
The train operation control system plays a key role in train operation safety and is regarded as the nerve center of the HSR system.
A standard has been set up for the train operation control system, which is known as European Train Control System (ETCS) \cite{Ai-2014, Diaz-2014}.
In order to make ETCS work better and create a digital standard for railway communications, a dedicated mobile communication system called the global system for mobile communications for railway (GSM-R) has been proposed by International Union of Railway (UIC) \cite{Zhong-2009}.

GSM-R has been widely used in HSR communications and can maintain a reliable communication link between the train and the ground.
However, GSM-R has some major shortcomings, such as insufficient capacity, low network utilization, and limited support for data services \cite{Sniady-2014}.
A broadband wireless communication system for HSR called long-term evolution for railway (LTE-R) has been presented in \cite{Masur-2009, Gao-2010} and determined in the 7th World Congress on High-Speed Rail \cite{Guan-2011}.
Broadband wireless communications can enhance the train operation by allowing an operation center to monitor real-time train-related data information, such as safety information and track diagnostic information \cite{Pareit-2012}.
In addition to the train control data transmission, LTE-R is also expected to provide passenger services such as Internet access and high-quality mobile video broadcasting \cite{Ai-2014, Barbu-2010, Diaz-2014}.
With the benefit of it, passengers can treat their journey as a seamless extension of their working or leisure environment.

To improve the capacity for wireless communications on the train, the future HSR communication networks are expected to be heterogeneous with a mixture of different networks and radio access technologies that can be simultaneously accessed by hundreds of users on the train \cite{Fokum-2010, Zhang-2010}.
For instance, the heterogeneous network architecture can be considered as a combination of satellite network, cellular network and wireless data network \cite{Pareit-2012}, where the advantage of each access network can be taken into consideration.
This architectural enhancement along with the advanced communication technologies such as multiple-input multiple-output (MIMO), orthogonal frequency-division multiplexing (OFDM) and radio over fiber (RoF), will provide high aggregate capacity and high spectral efficiency.
Nevertheless, the demand for HSR wireless communications is increasingly growing, for example, the estimated wireless communication requirement could be as high as 65 Mbps per train \cite{Liu-2012}.
To further relieve the contradiction between the increasing demand and limited bandwidth of HSR wireless communications, it is necessary to implement radio resource management (RRM) to improve resource utilization efficiency and ensure quality of service (QoS) requirements.
However, the traditional RRM methods (e.g., handover, power control and resource allocation) for common cellular communications may not be efficient in HSR wireless communications due to the following reasons, which are closely related to the characteristics of HSR scenario.
\begin{itemize}
  \item \textbf{High mobility}. The dramatic increase of train speed will cause frequent handover. Given a cell size of 1-2 km, a high-speed train of 350 km/h experiments one handover every 10-20 seconds \cite{Tian-2012}.
      To solve the frequent handover problem is one of the main functions of RRM in HSR wireless communications.
      Moreover, the fast relative motion between the ground and the train will lead to large Doppler shift and small coherence time. The maximum speed of HSR in China is currently 486 km/h, which induces a Doppler shift of 945 Hz at 2.1 GHz \cite{Liu-2012}. Thus, when implementing resource allocation for HSR communications, it is necessary to consider the fast-varying channel and inter-carrier interference (ICI) especially for OFDM technology.

  \item \textbf{Unique channel characteristics}. The moving train encounters diverse scenarios (e.g. cuttings, viaducts and tunnels) with different channel propagation characteristics \cite{Ai-2014, Ai-2012}, which causes that a single channel model could not depict features of HSR channels accurately. It brings a big challenge to RRM schemes, which should be adaptive to diverse scenarios along the rail with different channel models.
      Furthermore, the line-of-sight (LOS) component is much stronger than the multipath components especially in viaduct scenario, which implies that the propagation loss mainly depends on the distance between the base station (BS) and the train \cite{He-2014-VTC}.
      Since the distance varies with the train's position, the power control along the time has a large influence on system transmission performance \cite{Dong-2013-TVT}.

  \item \textbf{Heterogeneous QoS requirements}. Many types of services with heterogeneous QoS requirements and priorities will be supported on the train \cite{Pareit-2012}. The QoS performance in HSR wireless communications will be degraded because of high mobility and unique channel characteristics, especially for real-time services and critical core services that are critical for the train operation \cite{Lin-2010}. In order to improve system performance and satisfy heterogeneous QoS requirements, it is critical to design effective RRM schemes and resource optimization methods for multiple services transmission in HSR communications.
\end{itemize}

All these unique characteristics make it challenging to facilitate RRM design for HSR wireless communications.
Thus, a new look into the RRM problem in HSR communications is urgently required, where the network
architecture and unique characteristics of HSR scenario should be fully taken into consideration.
This paper systematically reviews and evaluates the various studies performed in the area of HSR communication systems.
There are also some survey papers discussing the research topics in HSR wireless communications, such as the network architectures \cite{Fokum-2010, Zhang-2010}, the handover schemes \cite{Zhou-2014}, and the channel characteristics and models \cite{Wang-2016, Chen-2015}.
Contrary to these surveys, this paper provides a comprehensible survey on HSR wireless communications from the perspective of RRM and optimization design.
Our goal is to present a detailed investigation and thorough discussion of current state-of-the-art RRM schemes for HSR wireless communications, as well as provide a better understanding of the RRM research challenges and open issues in HSR wireless communications.

\begin{figure}[!thb]
  \centering
  \includegraphics[scale=0.64]{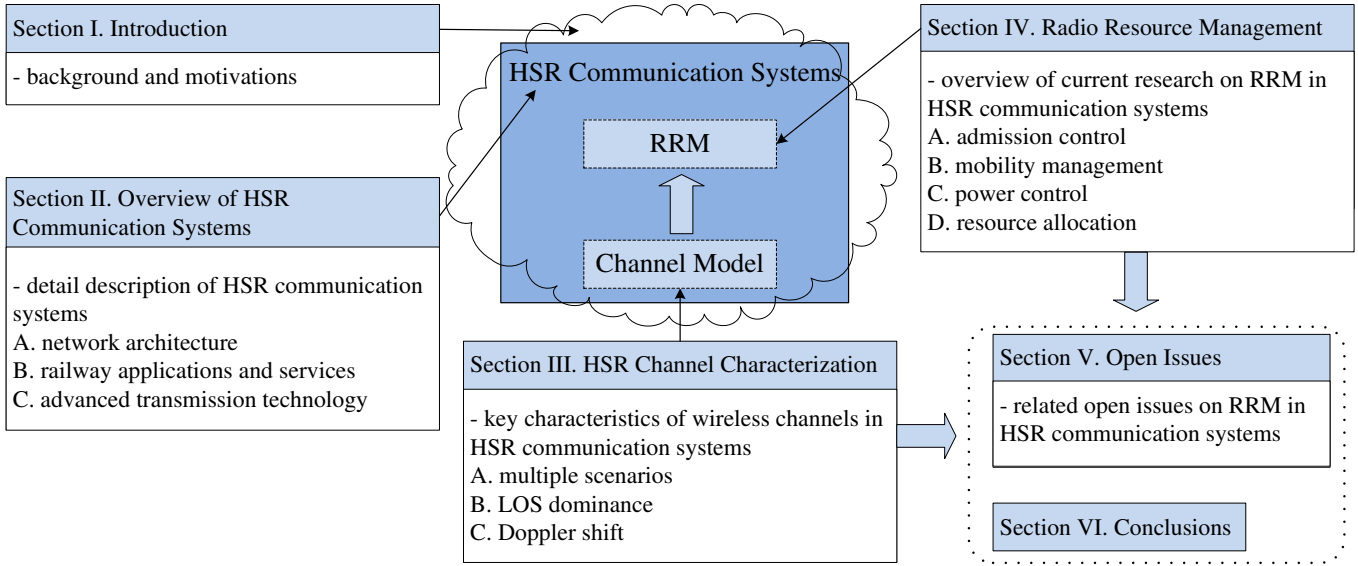}\\
  \caption{The structure of this survey paper}\label{Structure of Paper}
\end{figure}

The paper is organized based on the structure shown in Fig. \ref{Structure of Paper}.
Section II provides an overview of HSR communication systems, including potential network architecture, HSR applications and services, as well as advanced transmission technologies.
This is followed by a discussion on HSR channel characteristics in Section III, where a detailed analysis is carried out to understand the effects of channel characteristics on RRM design.
Section IV surveys the RRM schemes for HSR wireless communications from several aspects: admission control, mobility management, power control, and resource allocation.
The current challenges and open issues of RRM design for HSR wireless communications are given in Section V, prior to the conclusions in Section VI.

\section{Overview of HSR Communication Systems}
\subsection{Network Architecture}
\begin{figure}[!thb]
  \centering
  \includegraphics[scale=0.55]{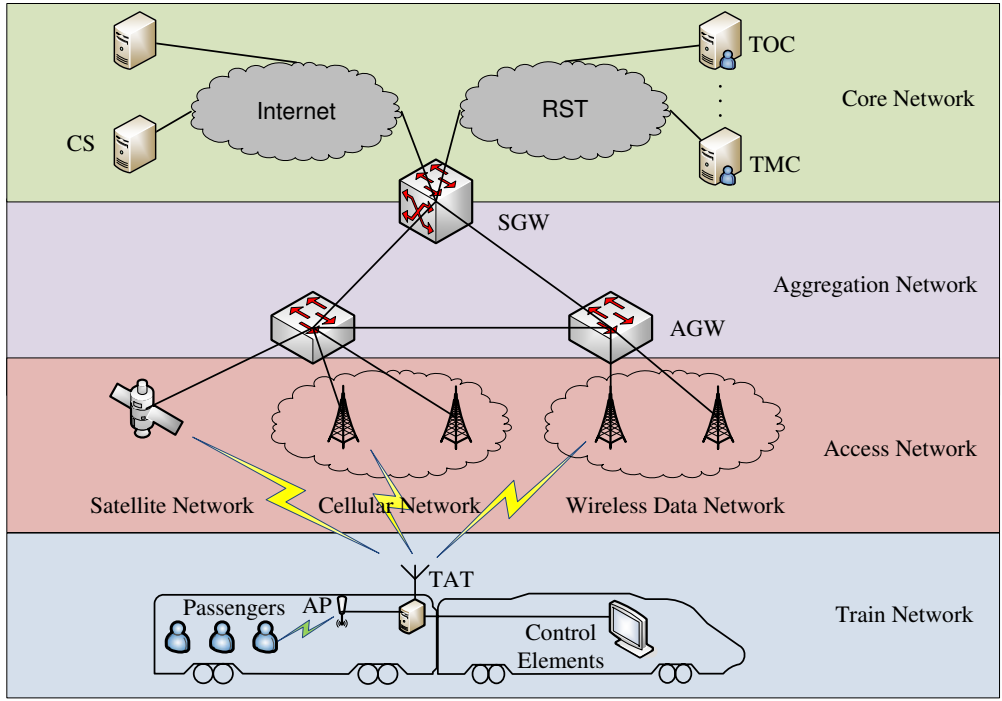}\\
  \caption{Network architecture for HSR communications}\label{Network Architecture}
\end{figure}

A proper network architecture is the basis of broadband wireless communications for high-speed trains \cite{Wang-2012}.
In this subsection, we provide a potential architecture to guide our discussion of broadband Internet access on trains.
A global overview of the network architecture for HSR communications is depicted in Fig. \ref{Network Architecture}.
This heterogeneous network architecture is responsible for the data transmission between the fast moving train and the service providers.
It is layered and consists of the core network, the aggregation network, the access network and the train network \cite{Pareit-2012, Greve-2005}.
All these four layers are briefly introduced as below.

\subsubsection{Core Network}
The core network is responsible for providing services and data processing in HSR communication systems.
There are two major actors in the core network, i.e., Internet and Railway Stakeholders (RST) \cite{Pareit-2012}.
The Internet can provide the multimedia services for passengers and the content servers (CSs) are deployed in order to offload data traffic \cite{Spagna-2013}, which have been applied in \cite{Liang-2012, Xu-2014-ACRA}.
The RST can have remote data access to devices on the trains, such as train operating company (TOC) and train maintainer company (TMC).
The former operates the trains for passenger or freight transport, while the latter is responsible for maintaining the trains.
In LTE-R system, the core network is the Evolved Packet Core (EPC) \cite{Gao-2010}.
The significant difference from the core network of GSM-R is that the EPC is an all-IP mobile core network.
This means that all services will be transmitted on the packet-switched domain.

\subsubsection{Aggregation Network}
The aggregation network lies between the access network and the core network, and it forwards data from the access network to the global Internet.
The aggregation network can use the following technologies for forwarding data: IEEE 802.11, Ethernet, ADSL, or optical fiber \cite{Fokum-2010}.
Two types of gateways in the architecture play a key role in data aggregation, named the access gateway (AGW) and service gateway (SGW).
The AGW serves as an interface between the access network and aggregation network.
It combines the data from the access networks and forwards that data to the SGW, which serves as an interface between the aggregation network and core network.

\subsubsection{Access Network}
The access network is close to the railway line, and it provides the last hop communication for the train.
Three kinds of wireless access technologies are used to provide wireless access for high-speed trains: satellite, cellular and dedicated wireless data networks.
Thus, this network architecture can be regarded as a heterogeneous network.
Table \ref{AccessNetwork} gives an overview of the different characteristics of these access technologies \cite{Pareit-2012, Zhang-2010, Fokum-2010}.
Specifically, it provides a comparison with respect to their capabilities of delivering broadband services to fast moving trains.
It can be seen that none of these technologies can meet all requirements: high data rate, low latency and fast handover.
Thus, to take full use of the advantages of different access technologies, a combination of some technologies is typically considered to be used in \cite{Lin-2002, Liang-2003, Kumar-2008}.
Meanwhile, the technical solutions and the choice of the right technology are well documented in the literature \cite{Verstrepen-2010, Fokum-2010}.

\begin{table}[!thb]
\renewcommand{\arraystretch}{1}
\caption{Comparison of different access networks}\label{AccessNetwork}
\centering
\begin{tabular}{|l|l|l|l|}
        \hline
         Parameter & Satellite network & Cellular network & Wireless data network\\
        \hline
        \multirow{2}{*}{Technologies}      & DVB-S, DVB-S2  &  2G, 3G, 4G  &  Wi-Fi, WiMAX \\
                                           & DVB-RCS        &  LTE         &  Flash-OFDM\\
        \hline
         Current converge & Very large (International) & Large (National) & Small (Limited)\\
        \hline
         Maximum train speed & Very high & High & Low-high\\
         \hline
         Supported data rate & High & Low-high & Very high\\
         \hline
         Transmission delay & High & Low-high & Very low\\
         \hline
         Handover performance & Good & Bad-good & Bad\\
        \hline
\end{tabular}
\end{table}

\subsubsection{Train Network}
Broadband wireless communications on the train are provided through the train access terminal (TAT).
This TAT connects to the access network using an antenna mounted on the outside of the train. It can combine different access technologies and can also intelligently select the best means of communication between the train and the access network \cite{Fokum-2010}.
The incoming signal from the TAT is then fed to the access point (AP) for passengers wireless access or control elements for the train controlling.

We have provided a detailed description of this heterogonous network architecture.
In conclusion, it has numerous advantages for HSR wireless communications
\begin{itemize}
  \item Since the TAT acts as a transmission relay, a two-hop communication link is established and penetration loss is avoided. In addition, the Doppler frequency shift and handover can be handled easily.
  \item This architecture can support multiple access technologies, thus the users don't need to switch or upgrade their equipments and just use their own radio access technologies.
  \item This architecture not only can provide multimedia services for passenger entertainment, but also can be used for railway signaling to improve railway operations.
\end{itemize}

\subsection{Railway Applications and Services}
\begin{figure}[!htb]
  \centering
  \includegraphics[scale=0.65]{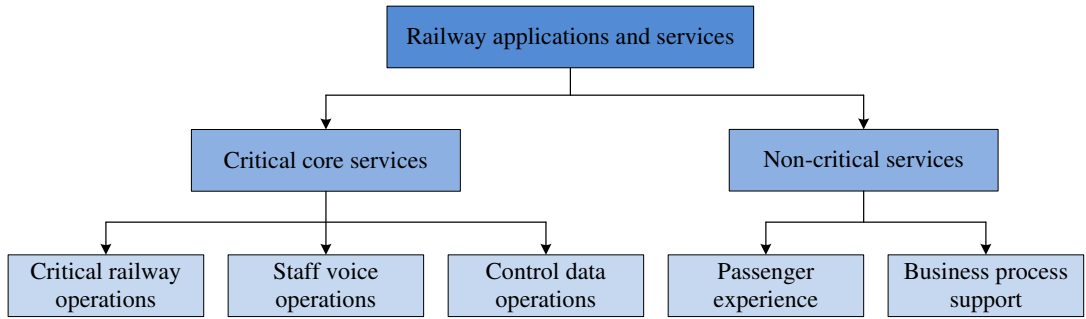}\\
  \caption{Railway services classification}\label{services classification}
\end{figure}

Current railway applications are focused on ensuring essential radio communications for train controlling, such as GSM-R and CBTC (Communication-Based Train Control).
The availability of high-speed broadband connections on the train opens up the possibility for new categories of services.
Fig. \ref{services classification} offers a comprehensive classification of the applications and services in HSR communications \cite{Masur-2009, Barbu-2010, Jaime-2013, Briso-Rodriguez-2014}.
Two main groups have been identified: critical core services and non-critical services.
Critical core services are usually referred to as mission critical services, including critical railway communications, train operational voice services and operational data applications.
The most common examples are staff voice communications, onboard closed circuit television (CCTV), and communication-based train control.
The non-critical services include passenger experience services and business process support services.
The passengers can have high-speed Internet access and personal onboard multimedia entertainment, which have been achieved by one of the European HSR operators \cite{Fokum-2010}.
Some business-related services can also be supported, such as remote diagnostics and location-based services \cite{Jaime-2013}.

The future HSR communication systems can address both critical and non-critical applications.
Obviously, the service requirements are significantly different from each other.
The main requirements for non-critical services include coverage, network capacity and cost requirements, while the requirements for critical services are mainly related to reliability, availability and prioritization.
Specifically, the work \cite{Pareit-2012} has identified the QoS requirements of these HSR services from the perspective of delay, jitter and loss.
The critical services need assurances for low delay and high reliability, while the non-critical services are constrained by available bandwidth \cite{Jaime-2013}.
Clearly, there exists a mismatch between the critical services and non-critical services, and thus it has a significant effect on the RRM and optimization design, which will be discussed in Section IV.

\subsection{Advanced Transmission Technology}
\begin{figure}[!htb]
  \centering
  \includegraphics[scale=0.8]{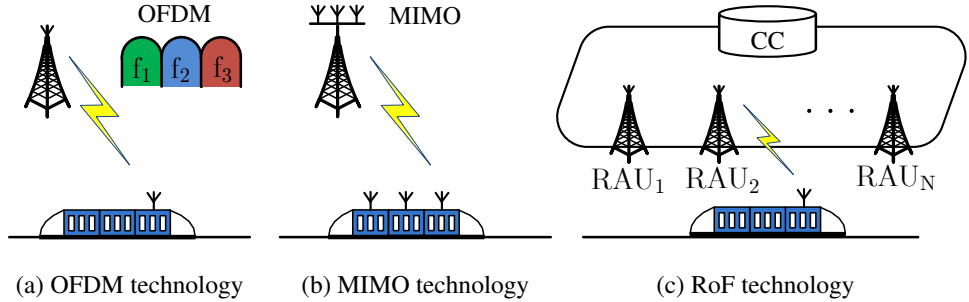}\\
  \caption{Advanced transmission technologies for HSR wireless communications}\label{AdvancedTechnology}
\end{figure}

The HSR wireless communications require high data rates to satisfy the demands of train operations and meet the communication requirements for passengers \cite{Zhou-2011, Briso-Rodriguez-2014}.
Advanced transmission technologies are desired to keep up with this unprecedented demand growth.
As shown in Fig. \ref{AdvancedTechnology}, OFDM, MIMO and RoF technologies can be efficiently utilized to improve the transmission capacity for HSR wireless communications.

\subsubsection{OFDM}
OFDM has become a dominant technology for the broadband wireless communications, forming the basis of all current WiFi standards and LTE. Its qualities include
\begin{itemize}
  \item The robustness to the frequency-selective multipath channel, avoiding intersymbol interference and leading to high spectral efficiency.
  \item The flexibility for resource allocation and management, since OFDM allows for dynamically assigning the subcarriers to multiple users and adaptively choosing the proper modulation and coding scheme.
  \item An excellent pairing for MIMO, since the spatial interference from multiantenna transmission can be handled at the subcarrier level, without adding intersymbol interference.
\end{itemize}

Given this impressive list of qualities, OFDM is the unquestionable frontrunner for future HSR communications \cite{Dai-2012, Qiu-2014}. However, some weak points would possibly become more pronounced in HSR environment.
First, Doppler shift caused by the high mobility destroys the orthogonality of the subcarriers, and results in ICI.
Since ICI significantly degrades system performance, it is not negligible when it comes to resource allocation.
Moreover, if OFDM technology is applied into future HSR communications, it is necessary and feasible to reduce the influence of ICI through channel estimation and equalization \cite{Yang-2012, Du-2012}.
Recently, a fundamentally distinct OFDM transmission scheme called time-frequency training OFDM is proposed in \cite{Dai-2012} for high speed mobile environments, and it can achieve high spectral efficiency as well as reliable performance.
Second, OFDM technology faces the peak-to-average power ratio (PAPR) problem, which will reduce the resolution of signal conversion and degrade the efficiency of the power amplifier.
For an actual power amplifier, a high PAPR sets up a design tradeoff between the linearity of the transmitted signal and the amplifier cost.
When applying OFDM technology into practical HSR communications, the PAPR problem can be largely overcome by using the pilot-assisted technique \cite{Popoola-2014} and signal precoding technique \cite{Hao-2010}.

\subsubsection{MIMO}
With the development of HSR communication systems, MIMO technology has become a hot research topic in train-to-ground communications \cite{Luo-2013, Wang-2011, WangChen-2012, Zhao-2012-ICC, Zhao-2012, Liu-2014-VTC}.
As we know, MIMO is one of the key technologies in 4G mobile communications, where the spatial multiplexing and transmit diversity can improve the signal quality and system spectral efficiency.
However, wireless channels are highly correlated with strong LOS due to few multipaths and scatterers in HSR environments.
The presence of the LOS component is regarded as a limiting factor for MIMO technology due to lower multiplexing gain compared with the prevalent Rayleigh fading channel \cite{Kang-2006}.
Thus, the conventional spatial multiplexing and diversity of MIMO cannot be directly applied into HSR environments.

Some advanced schemes have been proposed to take advantage of MIMO technology for HSR communications.
First, based on the 3-D modeling of the LOS MIMO channel, a multiple-group multiple-antenna (MGMA) scheme is proposed in \cite{Luo-2013} to increase the MIMO channel capacity, where the capacity gain can be achieved by adjusting the weights among MGMA arrays.
Second, beamforming is the most suitable transmission mode for the LOS scenario with strong channel correlation. With the help of real-time train position information, a beamforming platform and a distributed beamforming algorithm for HSR communications are respectively proposed in \cite{Wang-2011} and \cite{WangChen-2012} to improve the communication quality.
Finally, the combination of OFDM and MIMO technologies has been introduced into HSR communications, and it can further improve the spectrum efficiency \cite{Zhao-2012-ICC, Zhao-2012}.

Recently there have been increasing interests in massive MIMO systems, which employ tens or hundreds of antennas at the transmitter and receiver \cite{Rusek-2013}, and can potentially yield a significant improvement in system performance.
The massive MIMO technology is especially suitable for HSR communications due to the large size of the train. Meanwhile, with the linear topology of HSR cellular networks, multiple antennas can be employed at both BS and train.
More specifically, a two-antenna configuration with one antenna at train head and another at train tail is adopted by \cite{Tian-2012} and \cite{Zhang-2010-ROF} to enable a more efficient handover process.
The optimum design and performance analysis of HSR communications with massive MIMO are studied in \cite{Liu-2014-VTC}, where co-located antenna and distributed antenna layouts on the train are considered.

\subsubsection{RoF}
Cellular mobile networks have been well developed for train-to-ground wireless communications.
To provide broadband wireless communications for high speed trains, RoF technology has been extensively investigated as an appropriate solution \cite{Wang-2012, Zhou-2011, Zhang-2010-ROF, Chang-2011, Hou-2012, Lannoo-2007}.
Figure \ref{AdvancedTechnology}(c) shows an example of the RoF-based cellular mobile network.
Several remote access units (RAUs) are located along the rail and an optical ring interconnects them.
All RAUs within the same ring are under the supervision of a centralized controller (CC), where the signal processing is performed.
Compared with conventional cellular networks, it can be seen that the RoF scheme is more cost-effective, since it needs only a control equipment with $N$ low-cost RAUs to provide the same coverage as that of $N$ conventional BSs.
Moreover, these $N$ RAUs that are controlled by the same CC, can be used as repeaters for the same signal and there is no need for handover when the train moves from one RAU to the next. Thus, the RoF scheme can  reduce the number of handover, which is the main cause of dropped calls in HSR communications.
In addition, RoF technology can greatly improve the coverage efficiency of HSR cellular networks \cite{Zhang-2010-ROF, Han-2015}.

RoF technology has shown a promising future for broadband wireless communications in HSR environments.
Based on RoF technology, a distributed antenna system (DAS) has been implemented and experimented over railway communications \cite{Chang-2011}.
The targets of implementing DAS are to extend the service coverage for single BS and alleviate handover penalty.
The DAS system could also provide better signal quality and bandwidth efficiency \cite{Wang-2012, Hou-2012}.
By combining RoF technology with the ``moving cells" or ``moving frequencies" concept, the handover time can be reduced significantly \cite{Zhou-2011, Wang-2012}.
Besides, to support higher data rate and provide larger bandwidth, millimeter-wave bands (e.g. 60 GHz band) are suggested to be combined with RoF technology \cite{Lannoo-2007, Zhou-2011}.

\section{HSR Channel Characterization}
The knowledge of the wireless channel between the train and ground is vital to the optimal design and performance of HSR communication systems.
Different from traditional cellular scenarios, there are some specific characterizations in the HSR propagation environments such as multiple scenarios, LOS dominance and Doppler shift \cite{Liu-2012}.
Detailed information about these channel characterizations has been illustrated in \cite{Wang-2016} and \cite{Chen-2015}.
This paper focuses on analyzing the effects of these channel characterizations on RRM design.

\subsection{Multiple Scenarios}
\begin{figure}[!htb]
  \centering
  \includegraphics[scale=0.8]{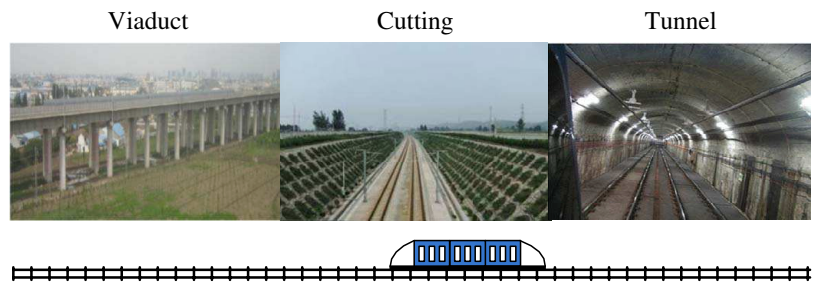}\\
  \caption{Multiple scenarios in HSR communications}\label{Scenario}
\end{figure}

HSR is usually situated on multiple different scenarios crossing a suburban or rural area.
The path loss and multipath effects are in disparity under various propagation environments, leading to the impossibility of accurate channel predictions under different propagation environments with the same channel model \cite{Ai-2014}.
Thus, radio wave propagation scene partitioning plays an important role in wireless channel modeling.
The HSR scenarios are partitioned into 12 types in \cite{Ai-2012}, where the detailed description about scene partitioning can be found.
The main HSR scenarios include viaduct, cutting and tunnel, as shown in Fig. \ref{Scenario}.

These special HSR scenarios have significant impacts on the propagation characteristics.
Viaduct is the most common scenario, where the railway track is placed on the viaduct surface to ensure its flatness.
The propagation channel in viaduct scenario is found to be greatly influenced by the physical environments, the directional BS antennas, and the viaduct height \cite{Liu-2012, He-2013}.
Cutting is used on uneven ground to help the train pass through large obstacles and ensure flatness of the rail lines. An important feature of the cutting is the rich reflection and scattering components caused by the steep walls on both sides of the railway tracks \cite{Ai-2012, HeR-2013}.
Tunnel is built through a mountain in HSR environments. For a long tunnel, several BSs are installed on the wall inside the tunnel. Due to the smooth wall and close structure, there are rich reflections and scattering components inside the tunnel, and they introduce the waveguide effect dominating the radio wave propagation \cite{Briso-2007, Guan-2012}.

Different propagation characteristics in these HSR scenarios bring several challenges to RRM design for HSR communications.
On one hand, the channel measurements and models for these HSR scenarios have been widely investigated in \cite{Liu-2012, Ai-2012, He-2013, HeR-2013, Guan-2012, Briso-2007}.
There are several types of channel models: empirical models, deterministic models, semi-deterministic models, and hybrid models \cite{Chu-2013}.
Generally speaking, empirical models are easy and fast to implement, but hard to reflect the practical propagation channel characteristics. In contrast, deterministic models can achieve higher accuracy than empirical models, but at the cost of high complexity.
Thus, when developing cross-layer optimization and RRM design, it is necessary to choose a suitable channel model based on the accuracy and complexity.
On the other hand, considering the complex environments along the rail, several propagation scenarios may exist in one communication cell \cite{Ai-2014}. When the high-speed train runs across these propagation scenarios, the channel characteristics change rapidly accordingly.
Thus, the cross-layer based RRM schemes for HSR communications should be adaptive to the rapid transition on channel propagation characteristics.

\subsection{LOS Dominance}
The newly-built HSR routes are distinguished from conventional ones, by gentler curves, shallower grooves and flatter tracks. Viaducts account for the vast majority of Chinese HSR (for example, 86.5\% elevated in Beijing-Shanghai HSR \cite{Luo-2013}). Combined with the high antennas, these yield an open space. Compared with urban and indoor environments, the LOS component is much stronger and other multipath echoes are encountered less frequently between the train and ground. This fact has been confirmed by engineering measurements \cite{Liu-2012, He-2011}. Thus, the channel is lightly scattered and only the LOS path is available at most time, which provides a convenient environment for HSR communications.

\begin{figure}[!htb]
  \centering
  \includegraphics[scale=0.8]{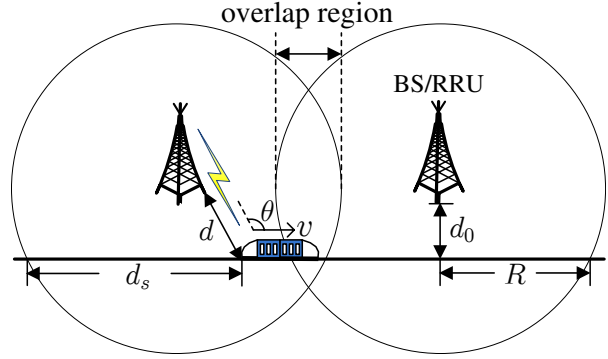}\\
  \caption{BS coverage model for HSR cellular communications}\label{LOS}
\end{figure}

The BS coverage model for HSR cellular communications is given in Fig. \ref{LOS}.
BSs are positioned along the railway, spacing each other by $2R$, which represents the service distance of one BS \cite{Zhang-2013}.
The train moves at a constant velocity $v$ along the rail. When the train locates at the position $d_s$ of current cell for $0 \leq d_s \leq 2R$, the distance between BS and train is $d = \sqrt{d_{0}^{2} + (d_s - R)^2}$, where $d_0$ is the distance between the BS and the rail line.
Based on the assumption of ignoring the altitude difference between BS antenna and train antenna, the propagation loss of the LOS path in free space is given by
\begin{equation}\label{propagation loss}
  L = 20\log_{10}\left[\frac{4\pi df}{c}\right],
\end{equation}
where $f$ and $c$ are the emitted frequency and velocity of light, respectively.
From \eqref{propagation loss}, we can see that the path loss $L$  is related with the distance $d$ and frequency $f$.

\begin{figure}[!htb]
  \centering
  \subfigure[]{
    \label{Pathloss} 
    \includegraphics[width=0.45\linewidth]{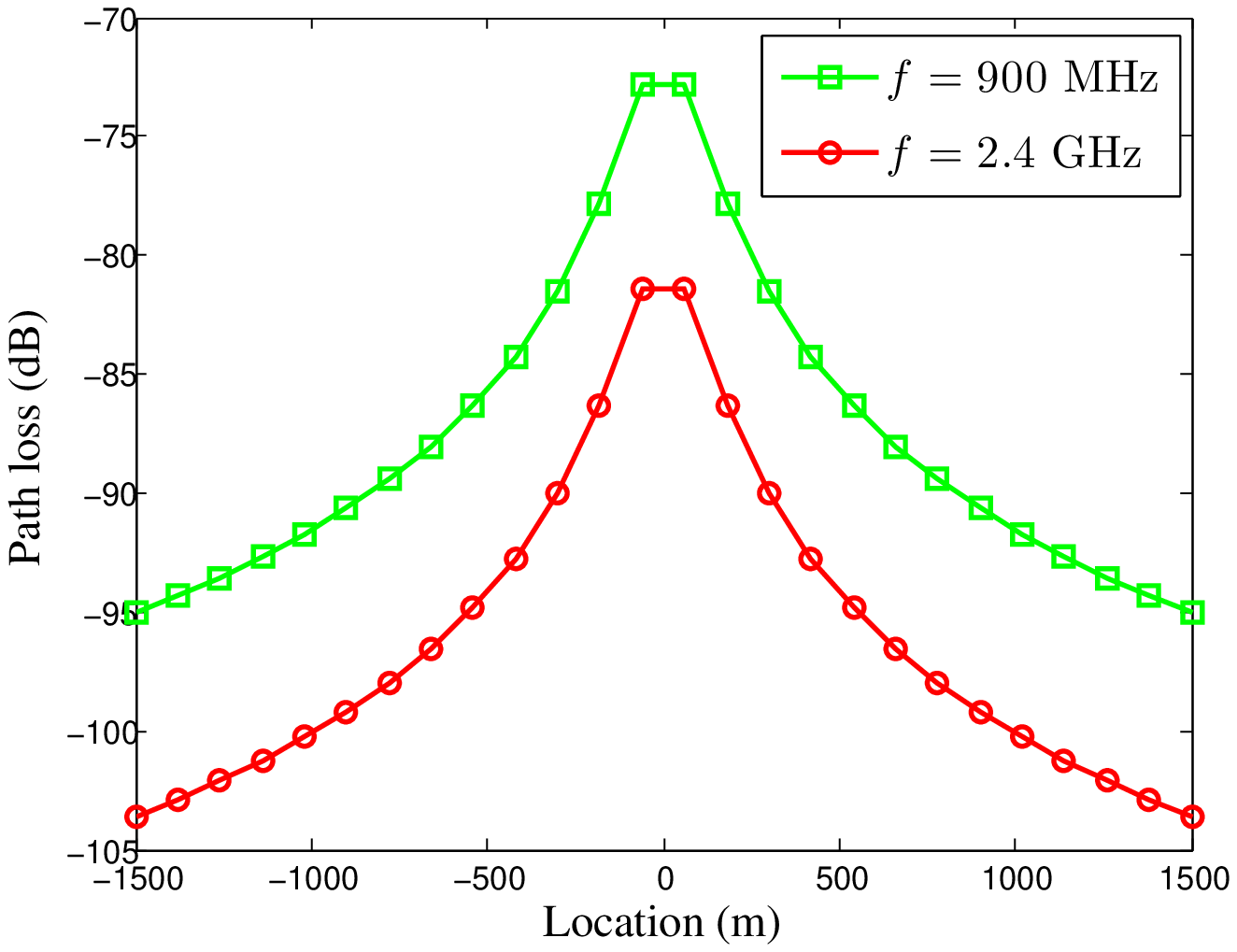}}
  \subfigure[]{
    \label{Dopplershift} 
    \includegraphics[width=0.45\linewidth]{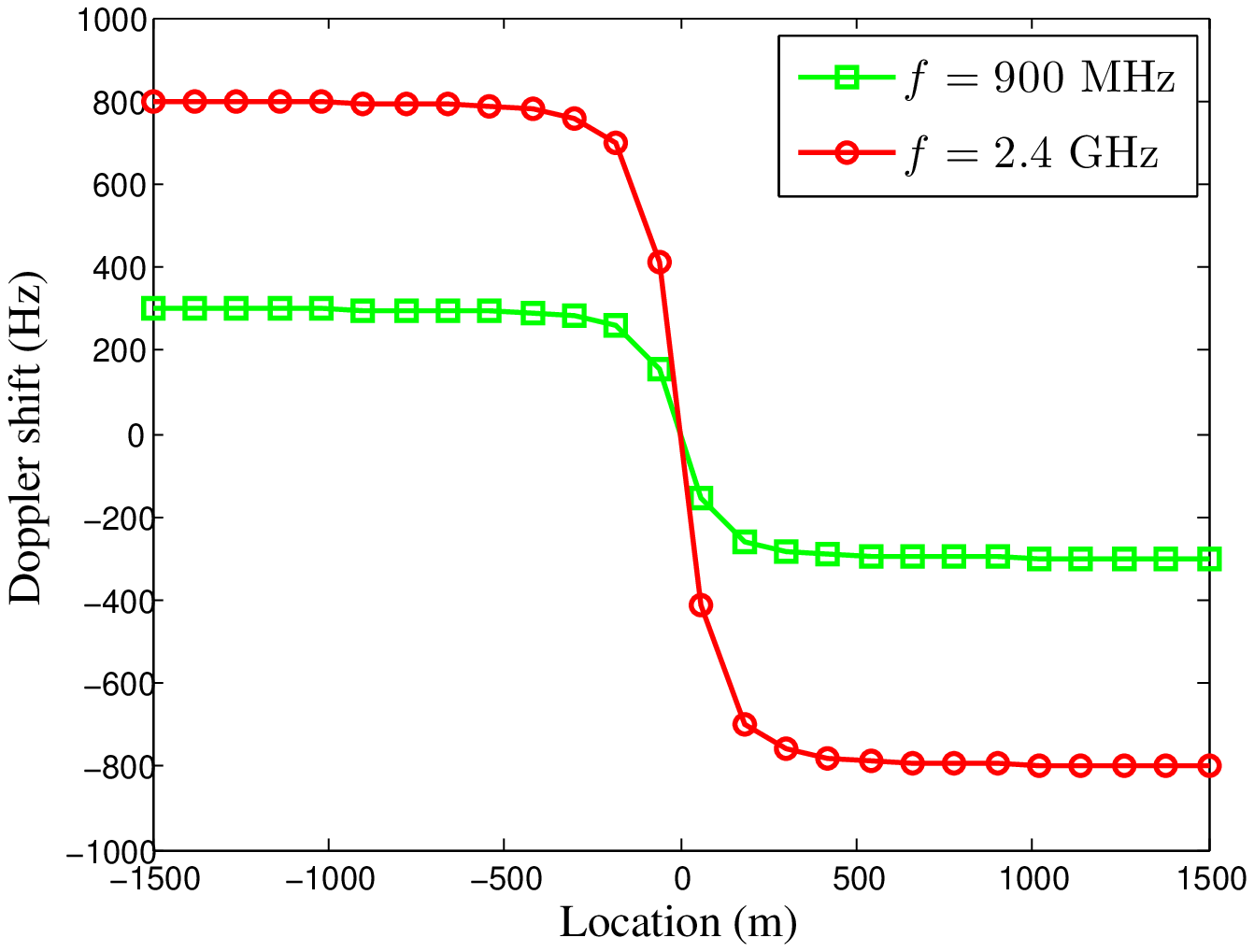}}
  \caption{Path loss and Doppler shift versus location at two frequencies: 900 MHz and 2.4 GHz. BS is located at 0, the cell radius $R$ is 1500 m, and the train moving speed $v$ is 100 m/s.}
  \label{Pathloss+Dopplershift} 
\end{figure}

Fig. \ref{Pathloss} shows the effects of both distance and frequency on path loss.
It is clearly shown that the path loss varies quickly with the train location no matter which frequency is used.
When the train moves towards the cell edge, the path loss turns larger and the corresponding channel condition gets worse. On the contrary, when the train moves towards the cell center, the path loss turns smaller and the corresponding channel condition turns better.
The periodic variation of channel conditions causes that the power control along the time has a large influence on the transmission performance. Thus, it is necessary and feasible to implement power control along the travel time.
It can be also observed from Fig. \ref{Pathloss}, the path loss at 2.4 GHz is larger than that at 900 MHz.
Note that 900 MHz and 2.4GHz are the working frequencies of GSM-R in China and LTE, respectively.
If higher frequency is chosen as the carrier frequency for future HSR communications, more transmit power will be consumed for pathloss compensation.
In addition, the path-loss value at cell edge and the selection of the carrier frequency will drive the network planning due to the different path losses at different distances and frequencies \cite{Fitzmaurice-2013}.

The path loss model \eqref{propagation loss} explains the effect of the periodic channel variation on network planning and RRM design.
To obtain reasonable network planning and optimal system performance, it requires a deep understanding of the path loss in HSR environments.
Based on the channel measurements in practical HSR environments, more accurate path loss models for different HSR scenarios have been widely investigated \cite{Liu-2012, He-2011, He-2014-VTC}.
Actually, the LOS path is a main component and the multipath can not be ignored in high-mobility scenarios.
Some channel models for high-mobility scenarios have been proposed in \cite{He-2013, HeR-2013, Guan-2012}, but they are very complex for the upper-layer protocol design and performance optimization.
To capture the characteristics of the time-varying wireless channel and simultaneously obtain mathematically tractable channel models, a finite-state Markov channel model is developed for HSR communications in \cite{Lin-2012}, and its accuracy is validated in \cite{Lin-2015}.

\subsection{Doppler Shift}
The high mobility causes the high Doppler shift and spread.
In HSR environments, the ratio of the energy in the LOS path to that in the multipath is relatively large, and the delay of multipath is relatively small.
As a result, the major influence on HSR wireless channels caused by high mobility is Doppler shift instead of Doppler spread \cite{Luo-2013}.
As shown in Fig. \ref{LOS}, when the train moves along the rail, Doppler shift $f_{\rho}$ can be calculated by
\begin{equation}\label{Doppler Shift}
  f_{\rho} = f_{d}\times \cos\theta,
\end{equation}
where $f_{d} = \frac{v}{c} \times f$ is the maximum Doppler frequency, $\theta$ is the angle between the train's forward direction and the line of sight from BS to the train.
Based on geometry information, from Fig. \ref{LOS}, we have $\cos\theta = \frac{R - d_s}{d}, ~ 0 \leq d_s \leq 2R$.
It can be seen that when the BS is placed far from the rail, i.e., $d_0 \gg R$, $f_{\rho}$ is relatively low since $\theta$ will be approximately $90^{\circ}$.
However, it will cause large path loss according to equation \eqref{propagation loss}.
Thus, there exists a tradeoff between path loss and Doppler shift when optimizing BS assignments.

Fig. \ref{Dopplershift} shows the Doppler shift along the rail for different carrier frequencies.
Several facts can be obtained as below.
\begin{itemize}
  \item $f_{\rho}$ is time-varying from the maximum positive value to maximum negative value as the train moves through a cell. This change process will repeat when the train moves along the rail.
  \item Severe Doppler shift exists in the communication process. The Doppler shift could be up to 800 Hz when the train travels at 100 m/s and the carrier frequency is 2.4 GHz.
  \item Although $f_{\rho}$ is very small as the train moves through BS, it will encounter a rapid Doppler transition.
  \item $f_{\rho}$ will jump from the maximum negative value to the maximum positive value, when the train moves into the overlap region between neighbor cells as depicted in Fig. \ref{LOS}.
\end{itemize}

For HSR wireless communications, severe Doppler shift and rapid Doppler transition issues need to be addressed before the practical application.
On one hand, severe Doppler shift would result in the difficulty of synchronization and bit error rate \cite{Yang-2012}.
However, it is worth noting that for most of the cell coverage, although the Doppler shift is large, its variation is so small that the Doppler shift can be estimated accurately and compensated easily according to accurate train's velocity and location information \cite{Du-2012, Li-2012-JWCN}.
On the other hand, rapid Doppler transition at the cell center greatly increases the difficulty of channel estimation and Doppler shift estimation, which will cause irrecoverable channel estimation error and severe ICI \cite{Zhang-2014, Liu-2012}.
The channel uncertainty and imperfection in the physical layer bring a big challenge to RRM design in the upper layers.
A reliable RRM scheme should be essentially robust to channel uncertainty and adaptive to the rapid Doppler transition.

\section{Radio Resource Management}
With growing demand for more and more QoS features and multi-service support in future HSR communication systems, RRM has become crucial and attracted great attention.
RRM is the process of developing decisions and taking actions to optimize the system resource utilization.
In particular, RRM consists of four key elements: admission control, mobility management, power control, and resource allocation \cite{Zander-2001}. Each element has a corresponding function with a common objective of achieving better system performance.
Compared with traditional cellular communications, supporting multi-service transmission under HSR scenarios introduces some new challenging issues to these RRM elements.
In this section, we provide a comprehensive state-of-the-art survey on RRM schemes for HSR wireless communications, with a focus on specific solutions for each element in detail.

\subsection{Admission Control}
Admission control is an essential tool for congestion control and QoS provisioning by restricting the access to network resources.
Generally, the admission control function has two considerations: the remaining network resources and the QoS guarantees.
In an admission control mechanism, a new access request can be accepted if there are adequate free resources to meet its QoS requirement without violating the committed QoS of the accepted accesses.
Thus, there is a fundamental tradeoff between the QoS level and the resource utilization.
To solve this tradeoff, admission control has been extensively studied in common communication networks, and different aspects of admission control design and performance analysis have been surveyed in \cite{Ahmed-2005}.
However, the admission control problem in HSR wireless communications is more sophisticated due to the following reasons.
\begin{itemize}
  \item High mobility brings some challenges to the implementation of admission control schemes, such as the fast-varying fading channel and frequent handover. The fast-varying characteristics of the wireless channel will cause channel estimation error, and further lead to admission control schemes inaccurate. For frequent handover, since available handover time is short and handover connections almost arrive in batches, it is critical to consider the handover connections when implementing admission control.
  \item Multiple services with heterogenous QoS requirements are supported on the train. The prioritization and fairness among different services should be taken into account in assigning admission control schemes. For example, higher priority should be given to safety-related services while the other services are delivered using the remaining network sources.
\end{itemize}
The above reasons make it difficult to directly apply the common admission control schemes into HSR communications.
The investigation on novel and proper admission control schemes is of great importance.
In the following, we provide an overview of the existing research on admission control for HSR wireless communications, including the classification, the detailed description, and the comparisons of different admission control schemes.

\subsubsection{Level-based Admission Control}
Based on the admission level, the admission control schemes can be classified into the call-level admission control and the packet-level admission control.
Traditional admission control schemes only consider call-level performance and are mainly designed for circuit-switched GSM-R system \cite{Zhao-2011, KimY-2013, Lattanzi-2010, Zhou-2014-AC, XuQ-2013-ICC, XuQ-2013, Zhao-2012-ICC}.
Since LET-R will become a packet-switched system, the packet-level features could be explored to improve the system performance \cite{Xu-2014-ACRA}.
The difference between the call-level admission control and packet-level admission control is as follows. At call level, each call is characterized by its arrival rate and holding time. If the system has enough resource, the new call request will be accepted, otherwise, the call will be rejected.
The packet-level features are characterized by the QoS profile that describes the packet arrival rate, packet queueing delay and packet loss ratio requirement.
For the elastic traffic transmission, the system simply drops the excess packets based on the system status and the minimum QoS guarantee.
To the best of our knowledge, the call-level admission control and packet-level admission control are investigated separately.
In the packet-switched LTE-R system, the packet-level dynamics are central to the call-level performance,
and thus both the packet level and call level should be considered in admission control schemes.

\subsubsection{Handover-based Admission Control}
Compared with the conventional mobile communications, handover is more frequent in HSR mobile communications. Thus, to reduce the handover failure probability and prevent the handover connections from being rejected, it is critical to consider the handover connections when implementing admission control.
Some existing papers have studied the admission control schemes associated with handover in high-speed mobile scenarios.
A spring model-based admission control scheme is proposed in \cite{Zhao-2011}, where every existing service is considered as a spring. For admitting handover services, bandwidth resource borrowing strategy is provided by compressing the spring as long as the lowest QoS can still be guaranteed.
Performance metrics such as dropping probability and blocking probability are analyzed based on the birth-death process.
In \cite{KimY-2013}, a mobility-aware call admission control algorithm is proposed in mobile hotspots, where a handover queue is involved to reduce the handover-call dropping probability.
The guard channels are dynamically assigned for handover calls depending on the vehicular mobility.
By means of Markov chains, the proposed algorithm is evaluated in terms of new-call blocking probability, handover-call dropping probability, handover-call waiting time in the queue, and channel utilization.
The literature \cite{Lattanzi-2010} proposes joint admission control and handover policies in an integrated satellite-terrestrial architecture, and the proposed policies aim to increase both the user satisfaction and the resource utilization.
Another handover-based admission control scheme is proposed in \cite{Zhou-2014-AC}, where a position-based factor is introduced to reserve more resources to accept handover calls. With the help of the handover location information, the resource reservation scheme divides the system resources for handover calls and new calls, respectively.

In the above schemes \cite{Zhao-2011, KimY-2013, Lattanzi-2010, Zhou-2014-AC}, the handover services and new services are considered when making admission control decisions.
Actually, the services can dynamically change their modulation and coding schemes (MCS) in HSR wireless communications.
When the adopted MCS changes from high spectrum efficiency to low spectrum efficiency, the occupied physical bandwidth will increase.
Then, besides handover services and new services, MCS changed service may be also dropped.
In \cite{XuQ-2013-ICC}, the main potential origins of service dropping are classified into two types: MCS changed service dropping and handover service dropping.
A cross-layer admission control scheme with adaptive resource reservation is proposed to reduce the service dropping probability, where the influences of MCS change and time-varying channels are taken into account.

\subsubsection{Priority-based Admission Control}
As described in subsection II.B, different types of services will be transmitted between the train and ground. Generally, different services have different priorities and bandwidth requirements. Thus, the service priority should be considered in admission control schemes for HSR wireless communications.
In \cite{XuQ-2013}, an admission control scheme with the complete-sharing resource allocation model is proposed for LTE-R system to maximize the number of admitted services while guaranteeing their related QoS. The proposed admission control scheme gives high priority to on-going services and guarantees the optimal bandwidth resource allocation.
\cite{Zhao-2012-ICC} proposes an effective admission control scheme for HSR communications with MIMO antennas, where both new services and handover services are considered. Handover services will be admitted first. The reason is that dropping an on-going service during the handover process will bring about more serious results than blocking a new service.
For different handover services to admit, voice has the highest priority while data has the lowest.
After all the handover services are processed, the new services will be considered in the same way as the handover services.
A joint optimal design of admission control and resource allocation is considered in \cite{Xu-2014-ACRA} for multimedia service delivery in HSR wireless networks. Different types of services are assigned to different utility functions that represent different service priorities. The considered problem aims at maximizing the total utility while stabilizing all transmission queues under the average power constraint. A threshold-based admission control scheme is proposed and the service with high priority gets large average throughput.

\begin{table}[!thb]
\renewcommand{\arraystretch}{1}
\caption{Summary of admission control schemes}\label{AC schemes}
\centering
\begin{tabular}{|l|l|l|l|l|l|}
        \hline
         Scheme & Brief description & Articles & Performance criteria & Comments\\
        \hline
        \multirow{2}{*}{Call-level} & The admission is conducted at the & \multirow{2}{*}{\cite{Zhao-2012-ICC, Zhao-2011, KimY-2013, Lattanzi-2010, Zhou-2014-AC, XuQ-2013-ICC, XuQ-2013}}& call
dropping probability; & Designed for circuit-switched \\
                                       & call level.&& call blocking probability & GSM-R system.\\

        \hline
        \multirow{2}{*}{Packet-level} & The admission is conducted at the & \multirow{2}{*}{\cite{Xu-2014-ACRA}}& packet queueing delay; & Designed for packet-switched\\
                                       & packet level. & &packet loss ratio  & LTE-R system.\\

        \hline
         \multirow{2}{*}{Handover-based} & The handover services are admitted & \multirow{2}{*}{\cite{Zhao-2011, KimY-2013, Lattanzi-2010,Zhou-2014-AC, XuQ-2013-ICC}} & handover service dropping; & Resource reservation for \\
                                            & based on different criterions. &  & MCS changed service dropping & handover services is required.\\
        \hline
         \multirow{2}{*}{Priority-based} & Different priorities are assigned to & \multirow{2}{*}{\cite{Xu-2014-ACRA, Zhao-2012-ICC, XuQ-2013}} & service blocking probability; & Handover and critical core \\
           & different types of services. &  & service delivery ratio &services have a high priority.\\
        \hline
\end{tabular}

\end{table}

Related admission control schemes mentioned in this subsection are summarized in Table \ref{AC schemes}.
Based on the survey on admission control research, we can make some conclusions as follows.
First, since LTE-R is designed as a packet-switched communication system, it is necessary to conduct admission control at packet level, which needs further investigations.
Second, it is serious to drop on-going services during the handover process and critical core services during the trip. Thus, the handover services and critical core services should have high priority to access the networks, where the resource reservation approach is also required.
Third, the handover-based admission control and priority-based admission control should be combined together so as to deal with handover services and new critical core services reasonably. At the same time, their advantages can be exploited jointly.
Finally, adaptive admission control schemes with low complexity are critical to suit the requirements such as frequent handover, quick decision-making duration and fast-varying wireless channel in HSR environments.

\subsection{Mobility Management}
Mobility management is one of the most important issues in HSR wireless networks, with the purpose of supporting seamless service transmission for the moving train.
Generally, mobility management includes two aspects, i.e., location management and handover management \cite{ZhuK-2011}.
In HSR wireless networks, location management has the functions of tracking and updating the location of mobile users on the train.
Handover management enables the train to seamlessly obtain transmission services after handover.
In particular, mobility management is essential to meet the following requirements for HSR wireless networks.
\begin{itemize}
  \item \emph{Realtime and accurate location update}: Realtime location update is necessary for both the train and the mobile users due to the high mobility of the train. The location information can be used to facilitate RRM design, such as handover and power control.
      Therefore, realtime and accurate location update is important to improve the overall performance of HSR wireless communications.
  \item \emph{Fast handover and seamless transmission}: Fast handover is a crucial requirement for HSR wireless communications, since the train spends only a short period of time moving through one cell. Seamless transmission is required especially for delay-sensitive applications (e.g., safety-related services).
  \item \emph{Group mobility}: A group of mobile users in a train need to execute handover simultaneously when entering a new cell. Moreover, location update requests from all mobile users will be handled along the trip.
      There is a need to consider the introduction of optimization solutions to reduce unnecessary signaling overheads and system loads in the group mobility \cite{Zhu-2013}.
\end{itemize}

\begin{figure}[!htb]
  \centering
  \includegraphics[scale=0.6]{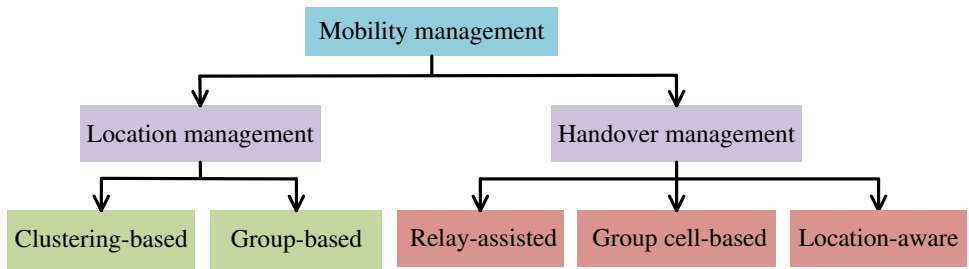}\\
  \caption{Taxonomy of mobility management for HSR wireless networks}\label{MobilityManagement}
\end{figure}

A simple taxonomy of mobility management for HSR wireless networks is shown in Fig. \ref{MobilityManagement}. In the following, various mobility management schemes are reviewed and discussed.

\subsubsection{Location Management}
There are a large number of mobile users in the same train. All mobile users will send the location update messages to the network at the same time. As a result, the signaling overhead increases and even causes the network congestion.
To reduce the cost of the location update and improve the system performance, the clustering-based location management (CLM) scheme and group-based location management (GLM) scheme are proposed in \cite{Wu-2012}, and they can ensure the mobile users to execute location update successfully.

The basic idea of CLM scheme is to divide the mobile users on the train into some clusters according to their identifications such as international mobile subscriber identity, and then the mobile users in each cluster perform location update in turns. Thus, it can reduce the frequency of location update and further decrease the network cost.
GLM scheme can be implemented based on the two-hop network architecture in Fig. \ref{Network Architecture}, where another device called group head should be equipped.
The group head is responsible for performing the integrated location update for the user group.
Specifically, the group location update is requested by the group head instead of each mobile user, and it may update the location information of all mobile users.
Apparently, the update efficiency is highly improved in both CLM scheme and GLM scheme.

Two features in HSR environments make it feasible to obtain accurate position information.
On one hand, since the train moves on a predetermined rail line and the velocity is relatively steady, the train trajectory information can be obtained in advance with high accuracy \cite{Liang-2012}.
On the other hand, many train positioning techniques are applied into HSR communication systems \cite{Pascoe-2009}. Some devices are installed on the train to determine its location on the rail, such as global positioning system (GPS) and odometer.
Thus, it is worthwhile to explore the potential of accurate position information for RRM design, such as handover management and power control in the sequel.

\subsubsection{Handover Management}
As shown in Fig. \ref{LOS}, when the train moves through the overlap region between two adjacent cells, the mobile users on the train will perform handover to the new cell.
Handover is one of the most challenging issues in HSR mobile communications, due to high handover frequency, group handover and QoS guarantees \cite{Tian-2012}.
The literature \cite{Zhou-2014} has presented an overview of the technical developments on handover in HSR mobile communications. Handover schemes are classified into seven categories according to the access technologies and network architectures. In this paper, we summarize the handover schemes from the perspective of the scenario characteristics in HSR mobile communications, and classify them into three groups, i.e., the relay-assisted, group cell-based and location-aware handover schemes.
In the following, the brief overview and comparison of these handover schemes are provided.

First, the relay-assisted handover schemes are based on the two-hop network architecture in Fig. \ref{Network Architecture}. The BSs only need to deal with the handover of relay rather than all mobile users, which leads to the reduction of drop-off rate.
A single relay on the train is involved in \cite{Huang-2012, Liu-2014} to facilitate handover process.
In particular, the authors in \cite{Huang-2012} modify the standard LTE handover scheme for the two-hop architecture, and bi-casting is introduced to achieve seamless handovers.
The work \cite{Liu-2014} proposes an effective handover scheme based on the distributed antenna system and relay-aided two-link architecture, where an antenna selection scheme with optimal power allocation is proposed to improve handover performance.
Some studies \cite{Tian-2012, ChenXJ-2013, Luo-2012, Lin-2014, Lee-2014-TWC, Yu-2015} describe the dual-link handover schemes based on the two-hop architecture. In these schemes, two relay stations are respectively installed on the front end and the rear end of the train. When the train is at the cell boundary, the front relay performs handover to the target BS, while the rear relay is still communicating with the source BS. In this way, the network connection can be maintained during the entire handover procedure. The results show that with the help of dual relays, soft handover can be effectively seamless, the handover interruption time is reduced, higher handover probability is guaranteed, and the system throughput is also improved.
Another relay-assisted handover scheme for LTE-advanced HSR networks is proposed in \cite{Pan-2015}, where a control relay is equipped in the front of the train, and several general relays are evenly distributed on the train. Mobile users are served by the general relays and the control relay performs the enhanced measurement procedure. The results show that the proposed scheme can effectively reduce handover time.
However, for the above relay-assisted handover schemes, since multiple relays need to be deployed on the train, the investment cost is high and lots of modifications on network procedures are required.

Second, group cell-based handover schemes have been extensively investigated recently. The group cell refers to a group of geographically adjacent cells. Handover only occurs between group cells and no handover is needed when the train moves within the group cell. Since the coverage region of one group cell is increased, it leads to the reduction of handover frequency.
RoF technology that has been briefly introduced in subsection II.C, can be regarded as a special case of group cell. The handover schemes for RoF technology are originated from the ``moving cell" concept, which is proposed in \cite{Gavrilovich-2001} to avoid frequent handovers in highway communications. This concept can be also applied into HSR communications, where the BSs physically move with the train. However, since the construction cost of a separate rail is high, the implementation of such a moving cell concept is difficult in practice \cite{Wang-2012}.
Alternatively, the moving cell concept can be replaced by the moving frequency concept \cite{Lannoo-2007, Zhou-2011}. The basic idea is to change the radio frequencies of the RAUs in a predefined order according to the train location. As a result, the frequencies of RAUs are moving with the train and the radio frequency used by the train will be unchanged. Although the handover is still needed when the train moves into the coverage area of another control center, the handover rate is reduced to a considerably low level.
As an extension of work \cite{Lannoo-2007}, the concept of ``moving extended cells" is proposed in \cite{Pleros-2009}. The main idea is to introduce a user-centric virtual extended cell consisting of a source cell and surrounding cells. As the train moves, a restructuring mechanism is applied to the virtual extended cell. In this way, the end-user is always surrounded by a certain number of grouped cells, which transmit concurrently the same data over the same radio frequency.
Another variation of moving cells is proposed based on a cell array in \cite{Karimi-2012}. The cell array smartly organizes a number of cells along the rail, together with a femtocell service that aggregates traffic demands within individual train cabins. Since the movement direction and speed of the train are generally known, the cell array can effectively predict the upcoming cells and enable a seamless handover.
Although the above group cell-based handover schemes can lead to the reduction of handover frequency, the construction cost will be high for dynamically organizing group cells.

Finally, in location-aware handover schemes \cite{Huang-2011, Karimi-2012, Aguado-2012, Cheng-2012, Fei-2012}, the handover process is facilitated with the help of train location information.
The basic idea in \cite{Huang-2011} is to divide the railways into several segments. When the train enters a predefined handover zone between two segments, the handover procedure can be automatically started.
In \cite{Aguado-2012}, the triggering condition of handover is adaptively decided according to the train's speed and position, when the train enters the coverage area of the target BS. To improve handover success probability, the mobile device can trigger handover earlier if the train moves at a high speed.
It is also pointed out in \cite{Karimi-2012} and \cite{Cheng-2012} that the forward-looking idea of using the predictable speed and location information can be applied to assist the handover process with the effective channel prediction.
Results from \cite{Fei-2012} show that position information can assist to reduce the handover delay and improve handover success probability under high mobility scenarios.
Two position-assisted fast handover schemes are proposed in \cite{Fei-2012}, with one making handover preparation before the handover procedure starts and another calculating the best target cell.
Although the train location information can be used to facilitate the handover process, the overheads are incurred in some aspects. First, trains and BSs have to exchange location information, and the trains have to carry GPS information in the measurement reports, which will cause standardization overhead.
Second, some works require the network operators to mutually plan the handover positions, which further increases the overhead on network planning.
Moreover, the location-based handover schemes may not be robust when GPS signal reception is poor or intermittent. For instance, when the train leaves a long tunnel, the GPS device on the train may need a long time to search satellites.

\begin{table}[!thb]
\renewcommand{\arraystretch}{1}
\caption{Comparison of handover schemes}\label{RelayHandover}
\centering
\begin{tabular}{|l|l|l|l|l|}
        \hline
         Schemes & Articles & Main ideas & Advantage & Disadvantage\\
        \hline
         \multirow{2}{*}{Relay-assisted} & \cite{Tian-2012} &The relay facilitates group handover& Group handover is avoided and & The investment cost is high\\
         & \cite{Huang-2012, Liu-2014, ChenXJ-2013, Luo-2012, Lin-2014, Lee-2014-TWC, Yu-2015} & procedures for mobile users onboard. &handover complexity is reduced. & due to deployment of relay.\\
        \hline
         \multirow{2}{*}{Group cell-based} & \cite{Lannoo-2007, Zhou-2011} & Increase the cell coverage and apply & The handover frequency can be & The cost will be high\\
          & \cite{Gavrilovich-2001, Pleros-2009, Karimi-2012} & various concepts of ``moving cell". & greatly reduced. & for constructing group cell.\\
        \hline
         \multirow{2}{*}{Location-aware} & \multirow{2}{*}{\cite{Karimi-2012, Huang-2011, Aguado-2012, Cheng-2012, Fei-2012}} & Train location information is applied& Handover process is facilitated & It incurs overheads and is  \\
          &  & to assist the handover process. &with better performance. &  influenced by location error.\\
         \hline
\end{tabular}
\end{table}

The comparison of different handover schemes is provided in Table \ref{RelayHandover}. Through the further analysis, we can observe that the handover performance can be improved by applying advanced technologies, such as relay and RoF technologies, at the cost of high system overhead.
To fully exploit the advantages of these schemes, the combination of them should be further investigated, such as \cite{Karimi-2012, Xia-2014}.
It is also necessary to study the handover scheme for heterogenous HSR networks with multiple access technologies \cite{Song-2014, Lin-2014}.
Finally, to provide seamless service transmission in HSR scenarios, an efficient handover scheme should be adaptive to fast-varying environment and robust to the location information error.

\subsection{Power Control}
\begin{figure}[!htb]
  \centering
  \subfigure[]{
    \label{PowerAllocation} 
    \includegraphics[width=0.48\linewidth]{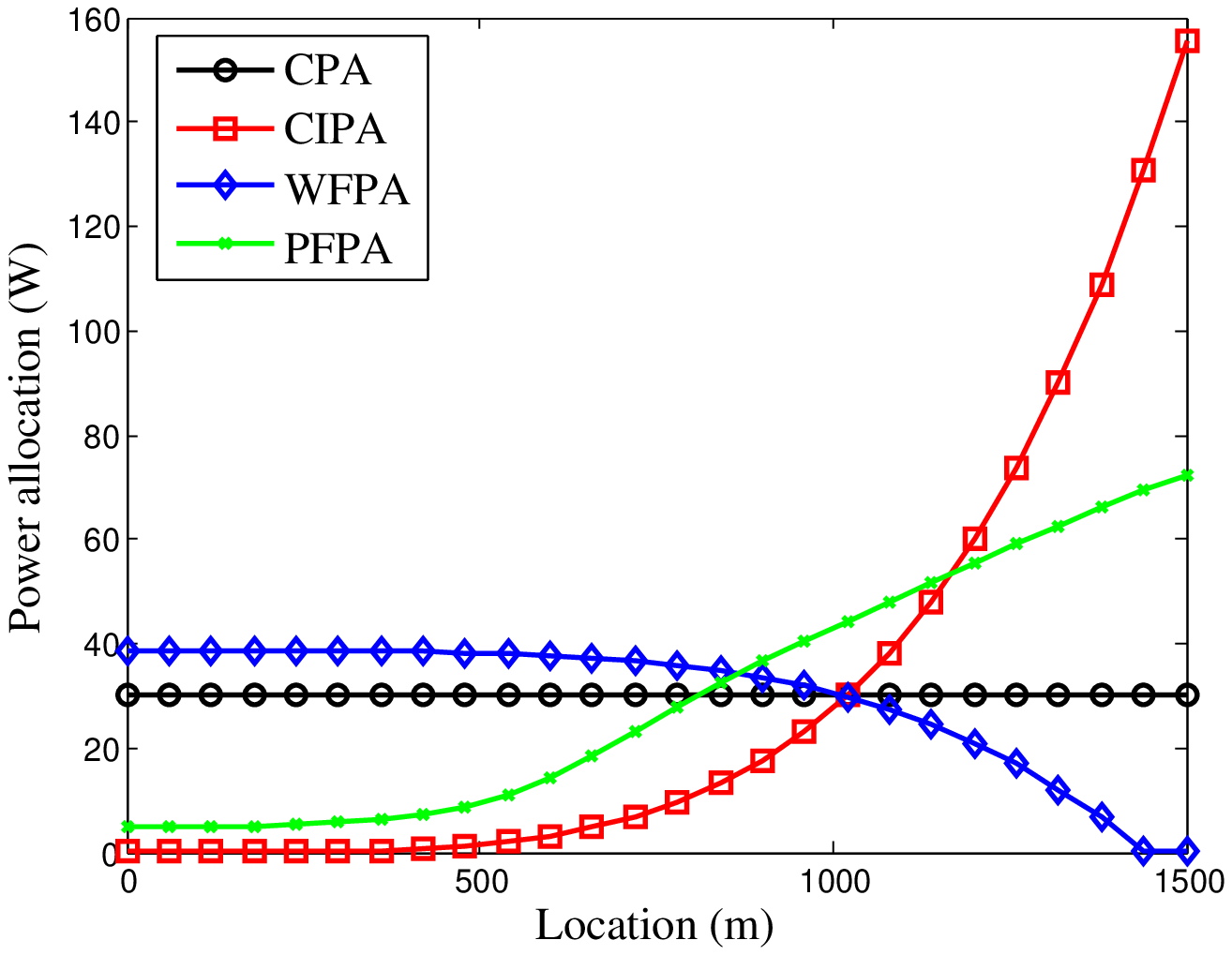}}
  \subfigure[]{
    \label{TransmissionRate} 
    \includegraphics[width=0.48\linewidth]{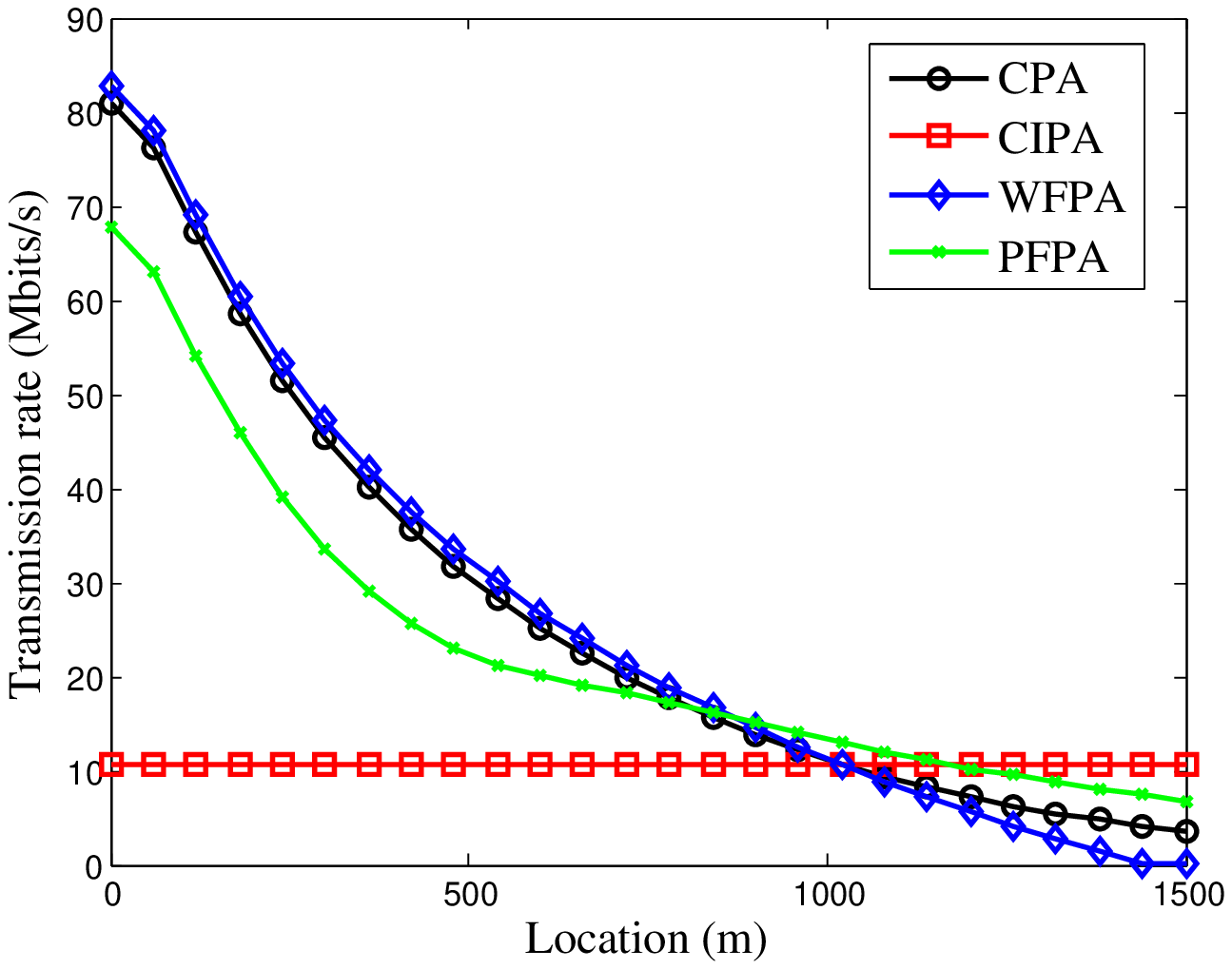}}
  \caption{Comparisons of four power allocation schemes \cite{Dong-2013}. BS is located at 0, the cell radius $R$ is 1500 m, and average power is 30 W.}
  \label{} 
\end{figure}

Compared with the conventional communication systems, there are three unique features in HSR communication systems \cite{Lin-2012}, i.e., the deterministic moving direction, relatively steady moving speed and the accurate train location information.
The data transmission rate is highly determined by the transmit power and the distance between BS and the train, thus these features make it necessary and feasible to implement power control along the time.
To achieve different optimization objectives under average power constraint, four power allocation schemes are proposed in \cite{Dong-2013-TVT}, i.e.,  constant power allocation (CPA), channel inversion power allocation (CIPA), water-filling power allocation (WFPA), and proportional fair power allocation (PFPA).
Fig. \ref{PowerAllocation} and \ref{TransmissionRate} provide the comparisons of the power allocation results and the corresponding transmission rate results for these four schemes.
The advantage and disadvantage of these schemes can be observed from these two figures.
For the sake of convenience in the engineering implementation, a constant power is allocated along the rail in CPA scheme while it ignores the variation of channel gain and results in the great unfairness in term of transmission rate.
In order to provide a stable transmission rate and achieve the best fairness along the rail, the CIPA scheme spends much power to compensate those bad channel states when the train is far from the BS.
Similar to the traditional water filling method, the WFPA scheme can maximum the total transmission rate within one BS, whereas the services will generally suffer from starvation when the train is near the cell edge.
In addition, the PFPA scheme can achieve a trade-off between the total transmission rate and the fairness along the time.
As an extension of \cite{Dong-2013}, the work\cite{Xu-2014-JWCN} investigates the utility-based resource allocation problem, which can jointly take into account power allocation along the time and packet allocation among the services.

The works \cite{Dong-2013} and \cite{Xu-2014-JWCN} pursue some system optimization objectives under the power constraint.
On the other hand, some works \cite{Thai-2013, Zhang-2015, Ma-2012} consider the power allocation problem from the perspective of energy efficiency, mainly focusing on how to match the data arrival process and the time-varying channel for HSR communications.
Specifically, \cite{Thai-2013} provides a novel method to minimize the total transmit power for data uplink transmission under a certain deadline constraint by exploiting the future channels.
In \cite{Zhang-2015}, authors study the optimal power allocation policy under given delay constraints in uplink transmission. It shows that there are two tradeoffs in the transmission model, one is between the average transmit power and the delay constraint, and the other is between the average transmit power and the train velocity.
Inspired by the unique spatial temporal characteristics of wireless channels along the rail, \cite{Ma-2012} presents a novel energy-efficient and rate-distortion optimized approach for uploading video streaming.
Although the studies in \cite{Dong-2013, Thai-2013, Zhang-2015, Ma-2012, Xu-2014-JWCN} are useful for the optimal design of HSR wireless communications, they only take account of the time-varying channel state while do not consider the dynamic characteristics of the service or packet arrivals, which causes that the above power allocation schemes are not practical.

Dynamic power control is necessary to improve the performance of HSR communication systems, where the transmit power should be adaptive to the time-varying channel and dynamic service arrival.
Considering the power constraint in HSR communications, the work \cite{Xu-2014-ACRA} investigates a joint admission control and resource allocation problem. A dynamic power control and resource allocation algorithm is proposed to maximize the system utility while stabilizing all transmission queues.
Different from \cite{Xu-2014-ACRA}, the work \cite{Xu-2014-DSO} studies the delay-aware multi-service transmission problem in HSR communication systems, with a focus on how to implement power control and resource allocation to guarantee the delay requirements under power constraint.

\subsection{Resource Allocation}
Resource allocation, as a critical part of RRM, plays an important role in enhancing the data transmission efficiency and improving the QoS performance in HSR communications. A great variety of resource allocation schemes have been proposed in the literature, aiming at sharing the limited network resources while satisfying the heterogeneous QoS requirements.
Due to the unique characteristics of HSR environments, the existing resource allocation schemes for general wireless networks cannot be directly applied to HSR wireless networks. Moreover, the dynamic characteristics such as time-varying wireless channels and random packet/service arrivals, should be incorporated into the resource allocation schemes.
In this subsection, the resource allocation schemes for HSR wireless communications are systematically surveyed from three aspects: interference-aware resource allocation, QoS-aware resource allocation and dynamic cross-layer resource allocation.

\subsubsection{Interference-Aware Resource Allocation}
Resource allocation is an effective way to reduce the effect of interference and further improve the communication performance.
Generally, the interference in HSR wireless communications mainly comes from two sources: ICI for OFDM technology and interference from two-hop links.
The ICI caused by Doppler shift, is obviously not negligible and may degrade the system performance.
Thus, the resource allocation problem in HSR communications with OFDM technology has attracted great research interest.
For example, \cite{Gong-2008} focuses on the joint subcarrier and power allocation problem with adaptive modulation and coding in high speed mobile environments.
The established optimization problem, which aims to minimize the overall transmit power while satisfying all the user requirements, is a non-linear programming. Considering that the exact expression of ICI term is complicated, the statistical mean value of the ICI power is utilized to simplify the complexity.
In addition, \cite{Zhao-2012} and \cite{Zhao-2013} study the multidimensional resource allocation problem for HSR downlink communications with OFDM technique and MIMO antennas. The objective is to maximize the throughput under the total transmit power constraint and ICI. In order to reduce computational complexity, the suboptimal solution is obtained by using quantum-behaved particle swarm optimization.

The above studies mainly focus on solving the resource allocation problems with ICI under one-hop architecture in HSR communications. Since the scarcity of spectrum nowadays, it is difficult to allocate dedicated spectrum for the two-hop links. Consequently, the spectrum reuse is inevitable and the resource sharing will lead to the inter-link interference. There are also some study works on resource allocation for the two-hop communications.
\cite{Qiu-2014} formulates a joint optimization problem of subcarrier allocation, subcarrier pairing and power allocation in two-hop links. Due to the complicated expression of ICI, the joint optimization problem is decomposed into several sub-problems and an iterative algorithm is proposed to solve the non-convex problem.
In \cite{XuShao-2013}, the resource allocation problem in two-hop links is formulated as a mixed integer nonlinear programming. Since the optimization problem is NP-hard, an alternative heuristic resource allocation scheme is proposed and then the optimal power is determined to maximize the system sum throughput.
In addition, the performance analysis of HSR wireless communications is studied in \cite{Zhang-2014} to better understand the effect of the inter-link interference.

\subsubsection{QoS-Aware Resource Allocation}
It is a natural requirement for HSR wireless networks to provide end-to-end QoS to the services, with the purpose of guaranteeing the safe operation of trains and improving the quality of passenger experience.
Data transmission reliability is critical in HSR communications, and is often evaluated by bit error rate (BER). With the speed increases, the communications will suffer from high BER, which may cause signaling error, retransmission and energy waste. Thus, the BER requirement should be considered for resource allocation design in HSR communications.
The work \cite{Zhou-2014-QoS} tries to solve a resource allocation optimization problem, which minimizes the total transmit power while considering BER that partially depends on the modulation and coding scheme.
For improving the goodput of HSR communications, a link adaptation scheme is proposed in \cite{XQS-2014-ICC} under the condition of guaranteeing prescribed BER target.
Multimedia entertainment is an important application in which a data stream often contains the packets with different BER requirements. For example, the video stream encoded with scalability contains the base layer packets with high BER requirement and enhancement layer packets with low BER requirement. To achieve efficient resource allocation, \cite{Zhu-2012} develops a resource allocation approach by considering multiple BER requirements for different types of packets in one data stream.
In order to simplify the complexity of resource allocation, a proper number of contiguous subcarriers are grouped into chunks and the spectrum is allocated chunk by chunk.

Service transmission delay is another key QoS parameter which directly affects perceived QoS of real-time service. The communication delay of train control services also affects the track utilization and speed profile of high-speed trains \cite{Lei-2016}.
HSR services such as critical core services have a high demand for transmission delay.
Thus, delay requirements should be fully taken into consideration when implementing resource allocation among the services.
To better describe delay requirements, an interval-based service request model is formulated in \cite{Liang-2012}, where each service has to be delivered within its given lifetime. The corresponding resource allocation problem aims to maximize the total weights of the fully completed requests and incomplete requests do not yield any revenue.
As an extension, \cite{Xu-2014-ICC} and \cite{Chen-2014} build more reasonable service delivery models for on-demand data packet transmission to high-speed trains, where the deadline constraints are built on each data packet rather than the whole service.
Compared with the deterministic deadline constraint case, \cite{Xu-2013} and \cite{Xu-2014-DADS} focus on the average delay constraint for the service transmission. The resource allocation problem is formulated as a constrained Markov decision process (MDP) and the corresponding online resource allocation algorithm is proposed.

\subsubsection{Cross-Layer Dynamic Resource Allocation}
Cross-layer design is a well-known approach to achieve QoS support. In cross-layer design, the challenges from the physical wireless channel and the QoS requirements are taken into account so that the resource allocation can be adapted to meet the service requirements for the given channels and network conditions.
It should also consider the dynamic characteristics in HSR communication systems, such as time-varying wireless channels and random service arrivals.
Thus, to enhance the efficiency of resource utilization and improve the QoS performance, it is necessary to implement dynamic resource allocation in a cross-layer way.

\begin{figure}[!htb]
  \centering
  \includegraphics[scale=0.75]{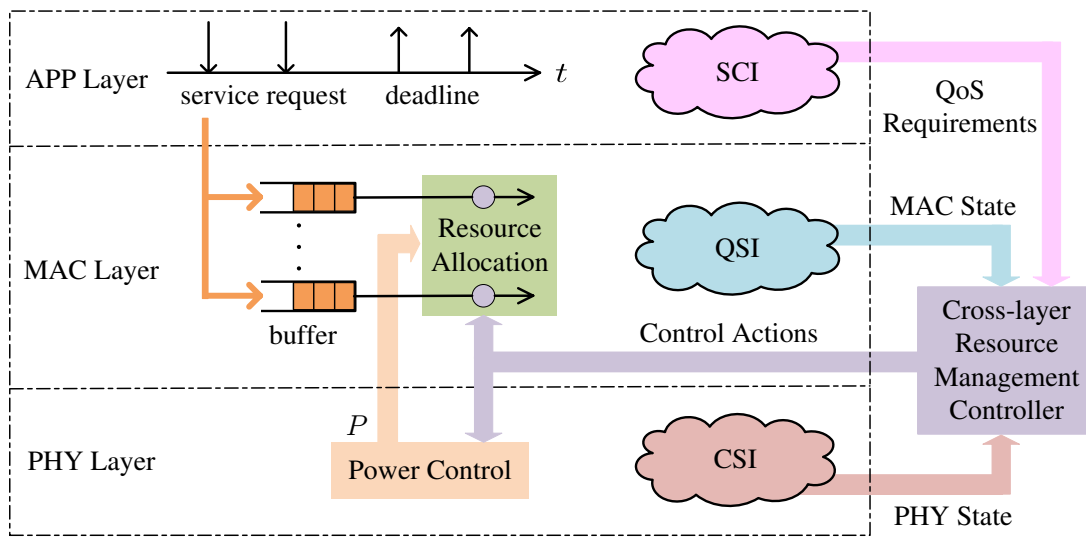}\\
  \caption{Illustration of cross-layer dynamic resource allocation framework}\label{DynamicCrossLayer}
\end{figure}

Fig. \ref{DynamicCrossLayer} presents an illustration of cross-layer dynamic resource allocation framework with respect to the application (APP) layer, medium access control (MAC) layer and physical (PHY) layer.
At the PHY layer, the channel state information (CSI) allows an observation of good transmission opportunity.
At the MAC layer, the queue state information (QSI) provides the urgency of data packets.
At the APP layer, service characteristics information (SCI) offers the service characteristics, e.g., packet arrival rate and rate-utility relationship.
The control actions, which include power control action and resource allocation action, should be decided dynamically based on the PHY layer CSI, the MAC layer QSI and the APP layer SCI.
Specifically, the power control action decides the wireless link capacity, i.e., the total resource allocated to the services. The resource allocation action decides how many resources should be allocated to each service.

Based on the above framework, \cite{Xu-2014-DADS} investigates the downlink resource allocation problem in relay-assisted HSR communication systems, taking account of bursty packet arrivals and delay performances. The considered problem is formulated as an infinite-horizon average cost constrained MDP, where the control actions depend on both the CSI and the QSI. The objective is to find a policy that minimizes the average end-to-end delay through control actions under the service delivery ratio constraints.
A joint admission control, power control and resource allocation problem is investigated in \cite{Xu-2014-ACRA}. A dynamic resource allocation algorithm is proposed to maximize the system utility while stabilizing all transmission queues.
In addition, \cite{Yang-2014} presents the resource allocation schemes for voice over Internet protocol (VoIP) traffic under HSR scenarios, in which the 2-state VoIP traffic model based on Markov chain is considered.
An adaptive resource allocation scheme is developed based on the idea of packet bundling to increase the spectral efficiency and reduce the outage probability.

\begin{table}[!thb]
\renewcommand{\arraystretch}{1}
\caption{Summary of resource allocation schemes}\label{ResourceAllocation}
\centering
\begin{tabular}{|l|l|l|l|l|}
    \hline
         Category & Articles & Problem characteristic & Challenges/Optimization method \\
    \hline
         & \multirow{2}{*}{\cite{Gong-2008}} &- Resource optimization with modulation &- Non-linear programming with great complexity\\
         &  &- Consider the effect of ICI &- Standard optimization package, e.g., CPLEX\\
       \cline{2-4}
         Interference-aware & \multirow{2}{*}{\cite{Zhao-2012, Zhao-2013}} &- Multidimensional resource allocation &- Computational complexity of joint optimization\\
         resource allocation &  &- Under MIMO and OFDM techniques &- Quadratic fitting/convex optimization\\
       \cline{2-4}
         & \multirow{2}{*}{\cite{Qiu-2014, XuShao-2013}} &- Resource allocation in two hop links &- The difficulty of non-convex optimization\\
         &  &- To handle two-hop interference &- Problem decomposition/convex optimization\\
    \hline
         & \multirow{2}{*}{\cite{Zhou-2014-QoS, XQS-2014-ICC, Zhu-2012}} &- Reliable transmission problem &- Simplified reliability expression with high accuracy\\
         & &- Consider BER requirements &- Partially observable MDP/convex optimization\\
       \cline{2-4}
         QoS-aware & \cite{Liang-2012} &- Reward maximization problem &- Low-complexity online scheduling design\\
         resource allocation & \cite{Xu-2014-ICC, Chen-2014} &- Deterministic deadline constraint &- Sequencing theory/Checker algorithm\\
       \cline{2-4}
          & \multirow{2}{*}{\cite{Xu-2013, Xu-2014-DADS}} &- Long term resource optimization &- How to handle state-space explosion problem\\
          &  &- Average delay constraint &- MDP/heuristic scheme\\
    \hline
         & \multirow{2}{*}{\cite{Xu-2014-ACRA}} &- Joint resource optimization &- Dynamic design and optimization \\
         Cross-layer dynamic &  &- Dynamic system model&- Stochastic network optimization theory\\
       \cline{2-4}
         resource allocation & \multirow{2}{*}{\cite{Xu-2014-DADS, Yang-2014}} &- Dynamic resource allocation problem &-  Cross-layer design and dynamic optimization \\
         &  &- Dynamic traffic arrival &- MDP/heuristic scheme\\
    \hline
\end{tabular}
\end{table}

A summary of the discussed resource allocation schemes for HSR communications is provided in Table \ref{ResourceAllocation}.
It can be seen that these existing schemes mainly focus on interference awareness, QoS requirements and cross-layer dynamic design, which are consistent with the characteristics of HSR communications.
Moreover, these three categories of resource allocation schemes correspond to different resource allocation problems.
For the interference-aware resource allocation, the optimization problems are mainly built on the ICI and inter-link interference. The complicated expression of the interference term makes the formulated problems nonconvex, and some heuristic algorithms are proposed to obtain the suboptimal solutions.
In the QoS-aware resource allocation, the formulated problems consider the reliability or delay requirements.
The reliability is often evaluated by BER, which acts as a constraint and is affected by the applied modulation and coding scheme. The delay requirements are typically represented by the deterministic deadline constraint or average delay constraint.
For the cross-layer dynamic resource allocation, the considered problems are dynamic optimization problems with cross-layer design. Stochastic optimization theory and MDP theory are used to effectively solve them.
Finally, we point out that the combination of different resource allocation schemes would be an interesting direction for future research.
For example, a practical and challenging problem is resource allocation for the delay-aware data transmission using OFDM technology in the two-hop wireless links.

\section{Challenges and Open Issues}
As we have seen from the previous sections, RRM services as an effective method to optimize the system resource utilization and provide QoS guarantees. Besides the existing research efforts, there are still some challenges and open research issues on RRM design, which will be discussed as follows.
In addition, we also recommend the readers refer to \cite{Moreno-2015} for more challenges and opportunities related to radio communications that railways will meet in both the near and far future.

\subsection{Location-Aware RRM}
Train positioning technique is one of the main techniques in HSR communication systems \cite{Pascoe-2009}.
How to use the position information to enhance the HSR communication performance has become a research trend.
However, only a limited number of studies have considered the train position information to facilitate the system design, such as position-based channel modeling \cite{Liu-2012}, position-assisted handover scheme \cite{Fei-2012}, position-based limited feedback scheme \cite{Yan-2015-TVT}, and location information-assisted opportunistic beamforming \cite{Cheng-2012-JWCN}.
In HSR scenarios, the channel condition mainly depends on the signal transmission distance, which results in the different channel conditions along the rail. Therefore, further research is needed to exploit the train position information to facilitate RRM design, by taking use of the future channel or signal prediction. For example, based on signal strength prediction, \cite{Ma-2012} proposes an energy-efficient mobile data uploading approach, and the energy consumption is heavily reduced.
Moreover, the seamless handover scheme can be further strengthened with the help of train position information and signal prediction.

Accurate position and speed measurements of the high-speed train are critical for the location-aware RRM in HSR communications. However, there exist measurement errors in the train positioning techniques, such as GPS and odometer \cite{Li-2012-JWCN}. Thus, studies on the effect of location uncertainty are needed to assess performance in practical HSR scenarios. Also, models that capture the variability of train location information, as well as methods that are robust to such inaccuracies are required.

\subsection{Cross-Layer-Based Joint RRM}
To satisfy different requirements stemming from various layers, an appropriate cross-layer model will be highly beneficial to improve the RRM performance in HSR wireless communications. These models should account for the parameters at the PHY layer (e.g., channel, power and modulation order), MAC layer (e.g., scheduling, queuing and automatic repeat request), network layer (e.g., routing and packet forwarding) and APP layer with different QoS requirements, and then optimally determine the resource management actions.
Such cross-layer models are able to manage the inherent tradeoffs at different layers in a comprehensive manner. More models and designs of both physical layer and higher layer for high mobility wireless communications have been discussed in \cite{Wu-2016}.
However, note that the incorporation of the constraints from different layers can further complicate the feasibility of RRM problems. In light of practicality, it is necessary to devise the low-complexity cross-layer techniques so that the close-to-optimal solutions can be identified while not severely compromising the overall system performance.

Considering the cross-layer model, most available studies in HSR wireless communications only focus on one single type of RRM, such as independent resource allocation and handover scheme. To achieve higher system performance, it is necessary and challenging to jointly optimize RRM schemes \cite{Xu-2014-ACRA}.
For example, a cross-layer design approach is proposed in \cite{Zhu-2011} to jointly optimize application-layer parameters and handover decisions to improve video transmission quality.
Further studies are needed to develop joint RRM schemes, such as jointly optimizing admission control, power control and resource allocation. Additionally, the tradeoff between the complexity and performance deserves future research.

\subsection{Energy-Efficient RRM for Media Delivery}
Providing multimedia services in HSR wireless communications will become a reality in the very near future, such as on-demand media services \cite{Liang-2012} and social network services \cite{Ai-2014-SNS}.
Some network architectures have been presented for multimedia service transmission over HSR communications, such as a mobile proxy architecture in \cite{Chuang-2014}. However, most of the existing studies only focus on RRM design for both voice and data services while multimedia services receive little attention. Thus, the effective RRM schemes for multimedia services delivery are critical with the purpose of high efficiency and green train communications \cite{Ma-2012,Huang-2014}.
Proper RRM design can lead to significant improvements in energy efficiency due to the strong dependence of power consumption on the distance between the train and base station. For instance, solely for a fairly constant rate, much power will be consumed to compensate for the fading effect when the train is far from the base station. From the perspective of energy efficiency, it is intuitive to transmit more data when the signal is strong and less data when the signal is weak. Therefore, energy-efficient media delivery problem should be further analyzed, and the related schemes such as power control and resource allocation are developed.
Furthermore, the QoS requirements should be also considered into RRM design for media delivery, such as playback delay and quality level.

\subsection{Robust RRM for Reliability Assurance}
Reliability requirements of HSR wireless communications pose great challenges to RRM design.
Generally, the HSR services especially critical core services have strict reliability requirements, such as very low bit error rate and packet loss probability. The reliable transmission of critical core services directly affects the safe operation of the train. Thus, how to design a robust RRM scheme for reliability assurance is an important issue.
However, most existing RRM schemes for HSR wireless communications are built on the perfect CSI assumption, while few investigations have been performed on the impact of channel estimation error in terms of resource allocation and beamforming design. Indeed, the channel estimation error that may result from the time-varying wireless channel, has a direct effect on the communication reliability, such as the bit-error-rate performance.
Further research is therefore needed to address a range of issues related to communication robustness, and improve the reliability of service transmission.
A typical example is the robust beamforming design problem for MIMO-based HSR wireless communications with imperfect CSI,
where the worst-case performance optimization will be exploited and the robust algorithms with low complexity are more attractive for practical implementation.

\subsection{Advanced RRM for Fifth Generation Communications}
The research for the fifth generation (5G) communications is now on its way. The European Union, the United States, Korea, China, and Japan have developed their organizations for 5G development. The HSR scenario has been recognized as one of the typical scenarios for 5G \cite{Ai-2014}. The related system designs such as network architecture and transmission technique have received much attention. Meanwhile, the advanced RRM for 5G HSR communications is urgently needed to further improve overall system performance and face the increasing demands.

To meet the emerging massive capacity demands in 5G communications, a control and data signaling decoupled architecture bas been presented for railway wireless communications \cite{Yan-2015}, in which the relatively important control plane is kept on high-quality lower frequency bands to handle mobility, while the corresponding user plane is moved to higher frequency bands to gain broader spectra. Since the control plane and user plane are physically separated, the RRM schemes for this novel architecture, such as handover and spectrum allocation schemes, are required to further investigated.
One of the 5G key techniques, massive MIMO technique, has been involved into HSR wireless communications \cite{Liu-2014-VTC}.
Wireless coverage based on massive MIMO for railway stations and train cars is proposed
to fulfill the requirement of high-data-rate and high spectrum efficiency \cite{Ai-2015}.
Further investigations on channel modeling and system-level modeling for HSR communications are still needed.
Additionally, the bandwidth shortage has motivated the exploration of the underutilized millimeter wave (mm-wave) frequency spectrum for future HSR broadband mobile communications \cite{Dat-2015, Choi-2014}.
However, mm-wave communications suffer from huge propagation loss, which indicates that it would be beneficial to investigate the use of power control to improve system performance.

\section{Conclusions}
Radio resource management is a powerful tool that enables high resource utilization and results in improved QoS performance. However, compared with common cellular communications, some characteristics in HSR wireless communications, such as high mobility, unique channel conditions and heterogenous QoS requirements, impose some challenges to the RRM design.
This leads to significant attention on the study of RRM under HSR scenarios.
In this paper, we have provided a literature survey on HSR wireless communications from a perspective of RRM design.
Firstly, we have presented an overview of the HSR communication systems with a detailed description of network architecture, railway applications and services as well as advanced transmission technologies.
Then the HSR channel models and characteristics are introduced, which are vital to the cross-layer RRM design and optimization.
Afterwards, we have surveyed the RRM schemes for HSR wireless communications, with an in-depth discussion on admission control, mobility management, power control and resource allocation.
Finally, challenges and open issues on RRM design of HSR wireless communications have been outlined.

\section*{Acknowledgment}
This work was supported by the Natural Science Foundation of China under Grant No. U1334202, the State Key Laboratory of Rail Traffic Control and Safety under Grant No. RCS2015ZT001, and the Fundamental Research Funds for the Central Universities under Grant No. 2015YJS032.

\bibliographystyle{IEEEtran}
\bibliography{Ref_HSR-Survey}
\end{document}